\documentstyle[preprint,aps,prd]{revtex}

\begin{document}

\draft
\preprint{}

\author{R.J.M. Covolan and A.V. Kisselev\thanks{On leave from Institute 
for High Energy Physics, 142284 Protvino, Russia} \\
\small Instituto de Fisica Gleb Wataghin, \\ 
\small UNICAMP 13083-970, Campinas, SP, Brasil}
\title{Charged Particle Multiplicity in Diffractive Deep Inelastic 
Scattering}
\date{}

\maketitle

\begin{abstract}
\baselineskip=0.4in
The recent data from H1 Collaboration on hadron multiplicity in 
diffractive DIS has been studied in the framework of perturbative QCD 
as a function of invariant diffractive mass. The formulas obtained 
explain the observed excess of particle  production in diffractive DIS 
relative to that in DIS and $e^+e^-$ annihilation. It is shown that the
results are sensitive to the quark--gluon structure of the  pomeron. 
Namely, the data say in favour of a super-hard gluon distribution at the 
initial scale.
\end{abstract}

\vspace{1cm}

\centerline{PACS numbers: 12.32.Bx, 12.40.Nn, 13.60.Hb}

\vspace{1cm}

\centerline{\bf To be published in Phys. Rev. D}

\newpage

\section{Introduction} 

One of the most remarkable and 
intriguing discoveries in hadron physics in the last years was the 
observation of hard interactions (namely, partonic activity) in 
diffractive events by the UA8 Collaboration \cite{ua8} at the CERN 
Collider. This discovery was inspired by a seminal paper by Ingelman 
and Schlein \cite{ingelman} in which such a possibility was foreseen as
a consequence of the pomeron having an internal structure and a 
quark/gluon content.

Experiments performed at HERA by ZEUS \cite{zeus_diff} and H1 
\cite{h1_diff}  collaborations provided further elements to this view 
through the obtainment of deep inelastic electron-proton scattering 
(DIS) events accompanied by large rapidity gaps adjacent to the proton 
beam direction. The presence of rapidity gaps in such events is 
interpreted as indicative that the internal structure of a colourless 
object carrying the vaccum quantum numbers, namely the pomeron, is 
being probed \cite{zeus_diff,h1_diff}. 

Further experimental evidences of hard diffraction have been reported by 
CDF and D0 collaborations in terms of diffractive production of W's and 
dijets \cite{cdf,d0}. 

Taking together, these evidences seem to indicate that the 
Ingelman-Schlein (IS) picture \cite{ingelman} is right at least {\em 
qualitatively}. However, a serious problem arises when one checks this 
model quantitatively. For instance, the model predictions 
systematically overestimate the diffractive production rates of jets 
and W's \cite{alvero}. There are reasons to believe that the 
discrepancy between predictions and data comes from the so-called {\em 
pomeron flux factor} \cite{dino}. In fact, a recent analysis 
\cite{Covolan} has shown that the pomeron structure function extracted 
from HERA data by using the IS approach is strongly dependent on which 
expression is used for the flux factor.

Recently, new data from H1 Collaboration~\cite{H1} on final hadronic 
states in diffractive deep inelastic process (DDIS) of the type
\begin{equation}
ep \rightarrow e'XY, \label{I}
\end{equation}
where $X$ and $Y$ are hadronic systems, have been presented. The system 
$X$ and $Y$ in (\ref{I}) are separated by the largest rapidity gap. $Y$ 
is the closest    to the proton beem ($M_Y < 1.6$ GeV) and squared 
momentum transfer at the $pY$ vertex, $t$, is limited to $|t| <  1 \ 
\mbox{\rm GeV}^2$, while $\langle  Q^2 \rangle$ is $21 \div 27 \ \mbox{\rm 
GeV}^2$~\cite{H1}. Both invariant masses,
$M_X$ and $M_Y$, are small compared to $W$, the centre-of-mass energy 
of the $\gamma^* p$ system.

In particular, charged hadron multiplicity have been studied 
in~\cite{H1} as a function of $M_X$ in the centre--of--mass system of 
$X$. The data obtained have been compared with the calculations of 
JETSET $e^+e^-$ (which is known to reproduce well the $e^+e^-$ data) 
and with the data on hadron multiplicities in deep inelastic scattering
(DIS)~\cite{EMC}. 
The interesting observations presented in~\cite{H1} are the following:
i) for $M_X > 10$ GeV $\langle n \rangle^{DDIS}(M_X, Q^2)$ is larger 
than charged hadron multiplicity in DIS, 
$\langle n \rangle^{DIS}(W, Q^2)$, at comparable values 
of $W = \langle  M_X \rangle$; ii) $\langle n \rangle^{DDIS}(M_X, Q^2)$ 
is also larger than charged hadron multiplicity in $e^+e^-$ annihilation, 
$\langle n \rangle^{e^+e^-}(s)$, taken at $\sqrt{s} = \langle  M_X \rangle$.

In the present paper we demonstrate that perturbative QCD is able to 
explain (qualitatively and quantitavely) the more rapid growth of 
hadron multiplicity in DDIS. So, no mechanisms besides gluon/quark jet 
emission with subsequent  jet fragmentation into hadrons are needed.
It is shown that the results on hadron multiplicity are very sensitive 
to the quark-gluon structure of the pomeron. An advantage of the method developed here is that it does not depend on the pomeron flux  factor.

The paper is organized as follows. In Section II, we present the QCD 
formalism for the description of final hadron states in ordinary DIS. 
In Section III, we extend this formalism to diffractive DIS and
apply it to describe the H1 data. Our main conclusions are summarized
in Section IV.

\section{Hadron Multiplicities in Hard Processes in 
Perturbative QCD} 

For the time being, we cannot describe a transition 
of quark and gluons into hadrons in the framework of QCD. Nevertheless, 
perturbative QCD enables one to calculate an energy dependence of 
characteristics of final hadrons produced in hard process
($e^+e^-$ annihilation into hadrons, DIS, Drell--Yan process etc.).
Mean hadron multiplicity, $\langle  n \rangle$, is one of the main 
features of final hadronic states. Hadron multiplicity in $e^+e^-$ 
annihilation has
been studied in a number of papers. The result looks like (see, for 
instance, \cite{Webber})
\begin{equation}
\langle n \rangle^{e^+e^-}(s) = a (\ln s)^b \exp \left( c \sqrt{\ln s} 
\right), \label{2.1}
\end{equation}
where $\sqrt{s}$ is a total c.m.s. energy of colliding leptons. As one 
can see, $\langle n \rangle^{e^+e^-}(s)$ rises more rapidly than $\ln s$, 
although more slowly than any power of $s$. Expression~(\ref{2.1}) 
describes well the available data.

Hadron multiplicity in DIS was calculated first in the framework of       
perturbative QCD in Refs.~\cite{Kisselev/Bassetto} (see 
also~\cite{Kisselev1}). It was shown that average multiplicity in 
lepton scattering off parton
\begin{equation}
eq(g) \rightarrow e'X, \label{II}
\end{equation}
$\langle n \rangle^{DIS}_{q/g}(W,Q^2)$, is related to $\langle n 
\rangle^{e^+e^-}(s)$, the average 
multiplicity in $e^+e^-$ annihilation, taken at $\sqrt{s} = W$ (up to 
small NLO corrections which descrease in $W$ and 
$Q^2$)~\cite{Kisselev/Bassetto}. 

In a case when quark distribution dominates (say, at $x \simeq 1$), the
relation between $\langle  n \rangle^{DIS}$ and $\langle  n 
\rangle^{e^+e^-}$ looks like 
(up to small NLO corrections which decrease in $W$ and 
$Q^2$)~\cite{Kisselev/Bassetto} 
\begin{equation}
\hat{\langle n \rangle}^{DIS}_q(W, Q^2) = \langle n \rangle^{e^+e^-}(W). 
\label{2.3}
\end{equation}

If we consider small $x$, we have to account for the gluon distribution
and the result is of the following form~\cite{Kisselev2} 
\begin{eqnarray}
\hat{\langle n \rangle}^{DIS}_g(W, Q^2) = \langle n \rangle^{e^+e^-}(W) 
\left[\,1 + 
 \frac{C_A}{2C_F} \varepsilon \left( 1 - \frac{3}{2} \,\varepsilon 
\right) \right], \label{2.4}
\end{eqnarray}
where $C_A = N_c$, $C_F = (N_c^2-1)/2N_c$ and $N_c$ is a number of 
colours.

The quantity $\varepsilon$ is defined via gluon distribution 
(see~\cite{Kisselev2} for details)
\begin{equation}
\varepsilon = \sqrt{\frac{\alpha_s(W^2)}{2 \pi 
C_A}}\frac{\partial}{\partial \xi} \ln D^g(\xi, x), \label{2.5}
\end{equation}
$\xi$ being the QCD--evolution parameter $\xi(W^2)=\int^{W^2} 
(dk^2/k^2) (\alpha_s(k^2)/2\pi)$.
Let us notice that $\varepsilon$~(\ref{2.5}) does not depend on a type 
of a target, and $\varepsilon$ is completely defined by the evolution 
of $D^g$ in $\xi$.  

Starting from the well--known expression for $D^g_g$ at small 
$x$~\cite{Dokshitzer},
\begin{equation}
D^g_g(\xi, x) = \frac{1}{\ln (1/x)} v \mbox{\rm I}_1(2v) \exp (-d \xi), 
\label{2.7}
\end{equation}
where $v=\sqrt{4N_c \xi \ln (1/x)}$, $d=(1/6) \left( 11N_c + 2N_f/N_c^2
 \right)$, $N_f$ is a number of flavours, we get
\begin{equation}
\varepsilon = \varepsilon (W^2, x) = \sqrt{\frac{\alpha_s(W^2)}{\pi}} 
\left[\sqrt{\frac{\ln (1/x)}{\xi (W^2)}} - \frac{d}{\sqrt{2N_c}} 
\right]. \label{2.8} 
\end{equation}
At $\ln (1/x) \gg 1$ (that is omitting the second term in RHS of 
~(\ref{2.8}))
we come to the asymptotical expression for $\varepsilon$ from 
Ref.~\cite{Kisselev2}.  

Using (\ref{2.3}) and (\ref{2.4}), we get in a general case
\begin{eqnarray}
\hat{\langle n \rangle}^{DIS} (W, Q^2) = \langle n \rangle^{e^+e^-}(W) 
\left[ \Delta +  \left( 1 + 
\frac{C_A}{2C_F} \varepsilon (W^2, x) \left( 1 - \frac{3}{2}  
\,\varepsilon (W^2, x) \right) \right) (1 - \Delta) \right], 
\label{2.9}
\end{eqnarray}
where $\Delta/(1 - \Delta)$ defines the quark/gluon ratio inside the 
nucleon.
As was shown in \cite{Kisselev2}, $\varepsilon$ (\ref{2.8}) is a 
leading correction to $\hat{\langle n \rangle}^{DIS}$ which rises with the 
decreasing of $x$.

All mentioned above is related to the multiple production of the 
hadrons in DIS of lepton off the parton~(\ref{II}). It is a subprocess 
of the process of lepton deep inelastic scattering off the nucleon
\begin{equation}
ep \rightarrow e'X. \label{III}
\end{equation}
According to Refs.~\cite{Kisselev1,Petrov,Kisselev3}, the corresponding
formula for mean hadron multiplicity in DIS looks like
\begin{equation}
\langle n \rangle^{DIS}(W, Q^2) = \hat{\langle n \rangle}^{DIS}(W_{eff}, 
Q^2) + n_0(M_0^2).
\label{2.11}
\end{equation}

The quantitiy $\hat{\langle n \rangle}^{DIS}$ has been defined above (see 
(\ref{2.9})). It depends on the effective energy, $W_{eff}$, available
for hadron production in the parton subprocess~(\ref{II}):
\begin{equation}
W_{eff}^2 = k W^2. \label{2.13}
\end{equation}

The average multiplicity of nucleon remnant fragments, $n_0$, is a 
function of its invariant mass $M_0^2 = (1 - z)(Q_0^2/z + M^2)$, where 
$M$ is a nucleon mass.

The efficiency factor $k$ in (\ref{2.13}) was estimated in 
\cite{Kisselev1} to be $\langle k \rangle \simeq 0.15 \div 0.20$ at present 
energies. That is why $\langle n \rangle^{DIS}(W, Q^2)$ is less than 
$\langle n \rangle^{e^+e^-}(W)$ and 
the growth of $\langle n \rangle^{DIS}(W, Q^2)$ in $W$ at fixed $Q^2$ 
is delayed in
comparison with that of $\langle n \rangle^{e^+e^-}(W)$.

\section{Average Hadron Multiplicity in Diffractive DIS}

Hadronic system $X$ in diffractive DIS (\ref{I}) is produced as a 
result of the virtual photon--pomeron interaction:
\begin{equation}
e{\rm I\! P} \rightarrow e'X. \label{IV}
\end{equation} 
The kinematical variables usually used to describe DDIS (in addition to
$W$ and $Q^2$) are
\begin{equation}
x_{{\rm I\! P}} \simeq \frac{M_X^2 + Q^2}{W^2 + Q^2} \quad \mbox{\rm 
and} \quad
\beta \simeq \frac{Q^2}{M_X^2 + Q^2}. \label{3.02}
\end{equation}

We start from the fact that pomeron has quark--gluon structure. It 
means that hadron production in process~(\ref{IV}) is similar to that 
in parton subprocess of DIS (\ref{II}). 

Recently a factorization theorem for DDIS has been 
proven~\cite{Collins}. Structure functions of DDIS coincide with 
structure functions of DIS and have the same dependence on a 
factorization scale. As a result, distribution functions of quark and 
gluons inside the pomeron, $D_{{\rm I\! P}}^{q(g)}$, must obey standard
DGLAP evolution equations at high $Q^2$~\cite{Altarelli}.

So, we can conclude that a hadron multiplicity of system $X$ in DDIS, 
$\langle n \rangle^{DDIS}$, is given  by expression (\ref{2.9}), in which $W$ 
is replaced by $M_X$, while $x$ is replaced by $\beta$. Namely, we get
\begin{eqnarray}
\langle n \rangle^{DDIS}(M_X, Q^2) &=& \langle n \rangle^{e^+e^-}(M_X) \times  
\nonumber \\ 
&\times& \left[\,\Delta_{{\rm I\! P}} + 
 \left(1 + \frac{C_A}{2C_F} \varepsilon (M_X^2, \beta) \left( 1 - 
\frac{3}{2} \,\varepsilon (M_X^2, \beta) \right) \right) (1 - 
\Delta_{{\rm I\! P}}) \right]. \label{3.1}
\end{eqnarray}
Here $\varepsilon (M_X^2, \beta)$ is given by formula (\ref{2.8}) and 
$\Delta_{{\rm I\! P}}/(1 - \Delta_{{\rm I\! P}})$ defines quark/gluon 
ratio inside the pomeron.
As both $D^q_{{\rm I\! P}}$ and $D^g_{{\rm I\! P}}$ obey DGLAP 
equations, the quantity $\varepsilon (M_X^2, \beta)$ and, 
consequently, hadronic multiplicity in DDIS are sensitive to the 
quark--gluon structure of the pomeron at starting scale $Q_0$.
 
As folllows from (\ref{2.11}) and (\ref{3.1}), 
\begin{equation}
\langle n \rangle^{DDIS}(M_X, Q^2) > \langle n \rangle^{DIS}(M_X, Q^2) 
\label{3.4} 
\end{equation}
because $\langle n \rangle^{DDIS}(M_X, Q^2)$ is defined by $\langle n 
\rangle^{e^+e^-}(M_X)$, while $\langle n \rangle^{DIS}(M_X, Q^2)$ is 
expressed via $\langle n \rangle^{e^+e^-}(M_X^{eff})$, the quantity
$M_X^{eff}$ being much smaller than $M_X$ by factor $k$ (see 
(\ref{2.13})).

For $M_X < 29$ GeV we have $M_X^{eff} < 5 \div 6$ GeV. It is known 
that in the region $W < 5 \div 6$ GeV function $\langle n 
\rangle^{e^+e^-}(W)$ rises 
logarithmically ($\sim \ln W$) that is slower than (\ref{2.1}). This 
results in more rapid growth of  $\langle n \rangle^{DDIS}(M_X, Q^2)$ 
in $M_X$ in comparison with $\langle n \rangle^{DIS}(M_X, Q^2)$.

We conclude from formula (\ref{3.1}) that  $\langle n 
\rangle^{DDIS}(M_X, Q^2)$ 
should exceed $\langle n \rangle^{e^+e^-}(M_X)$. Moreover, the ratio 
$\langle n \rangle^{DDIS}(M_X, Q^2)/\langle n \rangle^{e^+e^-}(M_X)$ 
has to grow in $M_X$ at fixed $Q^2$ 
(that corresponds to the rise in $1/\beta$ at fixed $Q^2$). All said 
above is in good qualitative agreement with the data from H1 
Collaboration~\cite{H1}.

In Fig.~1 we show the result of the fits of the H1 data by using our 
formula~(\ref{3.1}) (solid curve). In order to obtain this result, we 
proceeded as follows. For the quantity $\Delta_{{\rm I\! P}}$ which 
enters expression~(\ref{3.1}) we used
\begin{equation}
\Delta_{{\rm I\! P}}(\beta, Q^2) = \frac{D^{q}_{{\rm I\! P}}(\beta, 
Q^2)}{{D^{g}_{{\rm I\! P}}(\beta, Q^2)}+{D^{q}_{{\rm I\! P}}(\beta, 
Q^2)}}, \label{3.5}  
\end{equation}
where ${D^{g}_{{\rm I\! P}}}$ and ${D^{q}_{{\rm I\! P}}}$ are 
respectively the gluon and the  singlet quark distributions inside the 
pomeron that, as mentioned above, obey DGLAP evolution equations 
\cite{Altarelli}. For the distributions at the initial scale 
$Q_0^2=4\ GeV^2$ the following forms were employed:
\begin{eqnarray}
D^{q}_{{\rm I\! P}}(\beta, Q_0^2) &=& a_1\ \beta (1-\beta), \nonumber \\
D^{g}_{{\rm I\! P}}(\beta, Q_0^2) &=& b_1\ {\beta}^{b_2} (1-\beta)^{b_3}.
\label{gluon}
\end{eqnarray}
Thus, for the quark distribution we fixed an initial hard profile 
leaving free the normalization parameter, while for the the gluon 
distribution we have left all parameters free without imposing any sum 
rule. 
In order to perform the fit, other elements are needed. 
In Eq.~(\ref{3.1}), for $\langle n \rangle^{e^+e^-}(M_X)$ we have used 
the parametrization
\begin{equation}
\langle n \rangle^{e^+e^-}(M_X)= 2.392 + 0.024\ln(\frac{M^2_X}{M^2_0}) 
+ 0.913\ln^2(\frac{M^2_X}{M^2_0})
\end{equation}
with $M_0=1\ GeV$, taken from \cite{mirian}. Besides that, the 
experimental data shown in Fig.1 are given in terms of average values 
of $M_X, \beta,$ and $Q^2$ that do not obey strictly the kinematical 
relation (\ref{3.02}). In order to be faithful to data, we employed in 
our fit the parametrization 
\begin{equation}
\langle \beta \rangle = \frac{a}{(1 + b\ \langle M_X \rangle)^{c}},
\end{equation}
with $a = 1.63, b = 0.165\ GeV^{-1}$, and $c=2.202$.

The dotted (dashed) line in Fig.~1 corresponds to a pure quark (gluon) 
content of the pomeron. 
The solid line, which gives the fit of the H1 data, is obtained

with parameters $a_1=2.400,\ b_1=3.600,\ b_2=5.279,$ and $b_3=0.204$ 
in Eqs.~(\ref{gluon}), which are evolved to the respective $Q^2$-value 
corresponding to each experimental point. 

The initial distributions and their evolution with $Q^2$ are shown in 
Fig.~2. As can be seen, a super-hard profile (peaked at 
$\beta \sim 1$) was obtained for the gluon distribution at the initial 
scale, in qualitative agreeement with H1 analysis \cite{h1_novo}.

The $Q^2$--evolution of the normalized fractions of the pomeron 
momentum carried by  quarks and gluons is presented in Fig.~3. It is 
shown that quarks are slightly predominant over gluons at the initial 
scale, but that this relation reverses as $Q^2$ evolves. 
In this case, our results do not follow those obtained by H1 
Collaboration since their analysis \cite{H1} favour a fit in which 
predominate gluons carrying $\geq 80\%$ of pomeron momentum at the 
initial scale. It must be said, however, that the results of Fig.3 are 
consistent with limits established in other experiments (see, for 
instance, \cite{cdf}).

\section{Conclusion}

We have presented in this paper a description of the hadron 
multiplicity obtained in diffractive DIS and recently reported by the 
H1 Collaboration \cite{H1}. This description was derived from a 
formalism previously developed for ordinary DIS within the framework 
of the perturbative QCD. 

The formula obtained enables us to explain the observed excess of 
hadron multiplicity in diffractive DIS, $\langle n 
\rangle^{DDIS}(M_X, Q^2)$, 
relative to those in DIS and $e^+e^-$ annihilation taken at $W = \langle 
M_X \rangle$ and $\sqrt{s} = 
\langle M_X \rangle$, correspondingly. The more rapid growth of 
$\langle n \rangle^{DDIS}(M_X, Q^2)$
is also understood.

It was shown (Fig.1) that neither a pure quark nor a pure gluon 
content of the pomeron satisfy the data, but that a mixing of both 
components in approximately equal shares is able to provide a good 
description of the experimental results. 
The pomeron structure function that comes out from the present 
analysis consists of a hard distribution for the quark singlet and 
a super-hard distribution for gluons at the initial scale of 
evolution, in agreement with what has been reported lately in the 
literature \cite{h1_novo}.

This result is remarkably good if one considers that it was obtained 
from a very small data set (only 7 data points), covering a limited 
$\beta$ range ($0.03 < \beta < 0.43$) for an equally restricted 
$Q^2$-range ($22 < Q^2 < 27\ GeV^2$).

Concluding, the most important result presented here is the theoretical
framework summarized by Eq.(\ref{3.1}), that has made it possible to 
explain anomalous H1 data on hadron multiplicity in DDIS. It also 
enabled us to extract information about the pomeron structure function 
from such a limited data set, without the usual complications and 
ambiguities with flux factor normalization and $x_{{\rm I\! P}}$-dependence. 

This conclusion points out to the need of more and more precise DDIS 
multiplicity data, taken at extended kinematical ranges. Such a 
possibility would improve a lot the analytical capacity of the scheme 
presented here. 

\section*{Acknowledgements}
We would like to thank the Brazilian governmental agencies CNPq and 
FAPESP for financial support.

\newpage

\section*{Figure Captions}

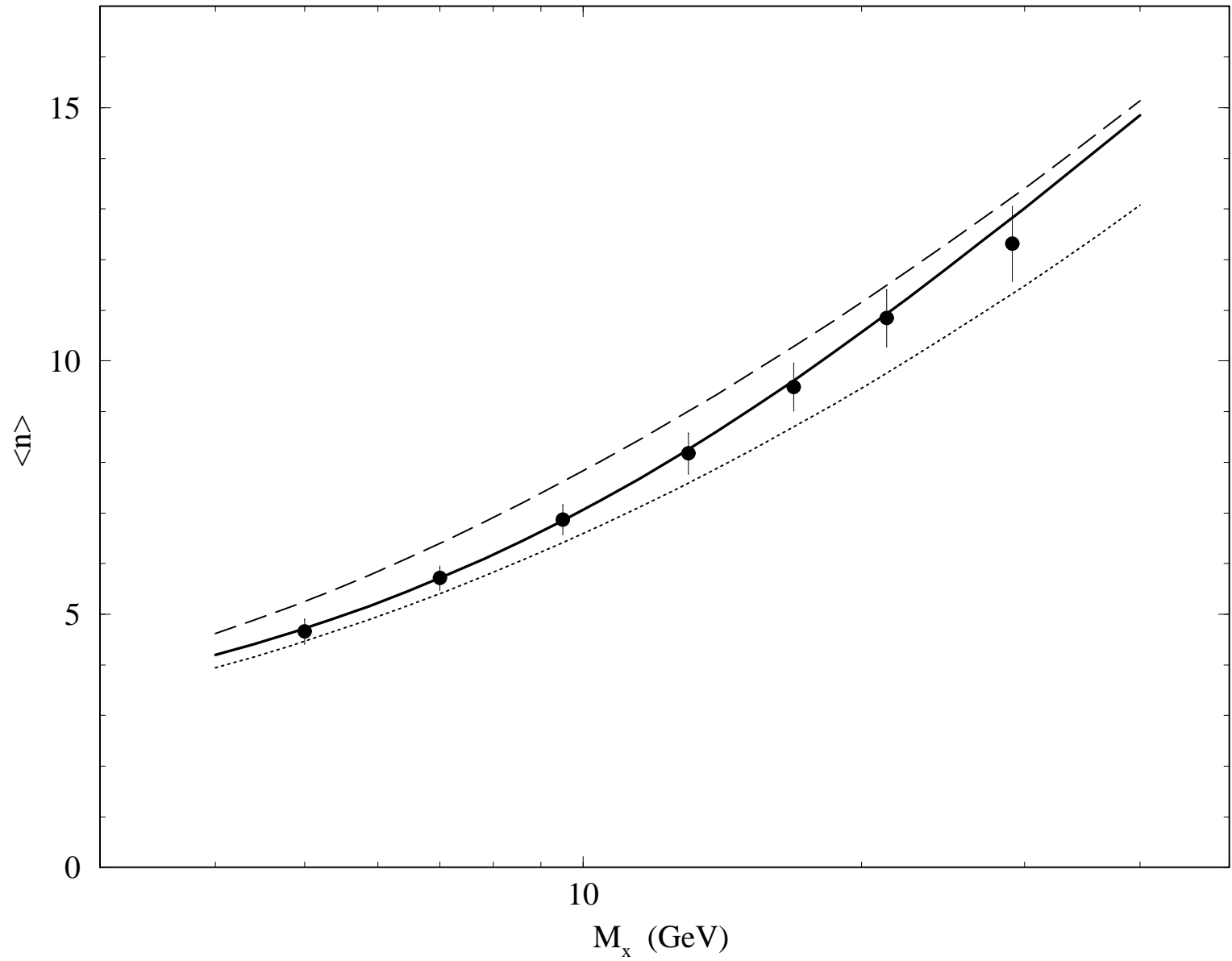
Figure 1: Description of the average multiplicity of charged hadrons 
produced in diffractive DIS experiments. Data obtained by the H1 
Collaboration \cite{H1}. The theoretical results are obtained from 
Eq.(\ref{3.1}) with different assumptions for the quark/gluon content 
of the pomeron: pure quark (dotted curve), pure gluon (dashed curve) 
and both components (solid curve).

\vspace{1cm}

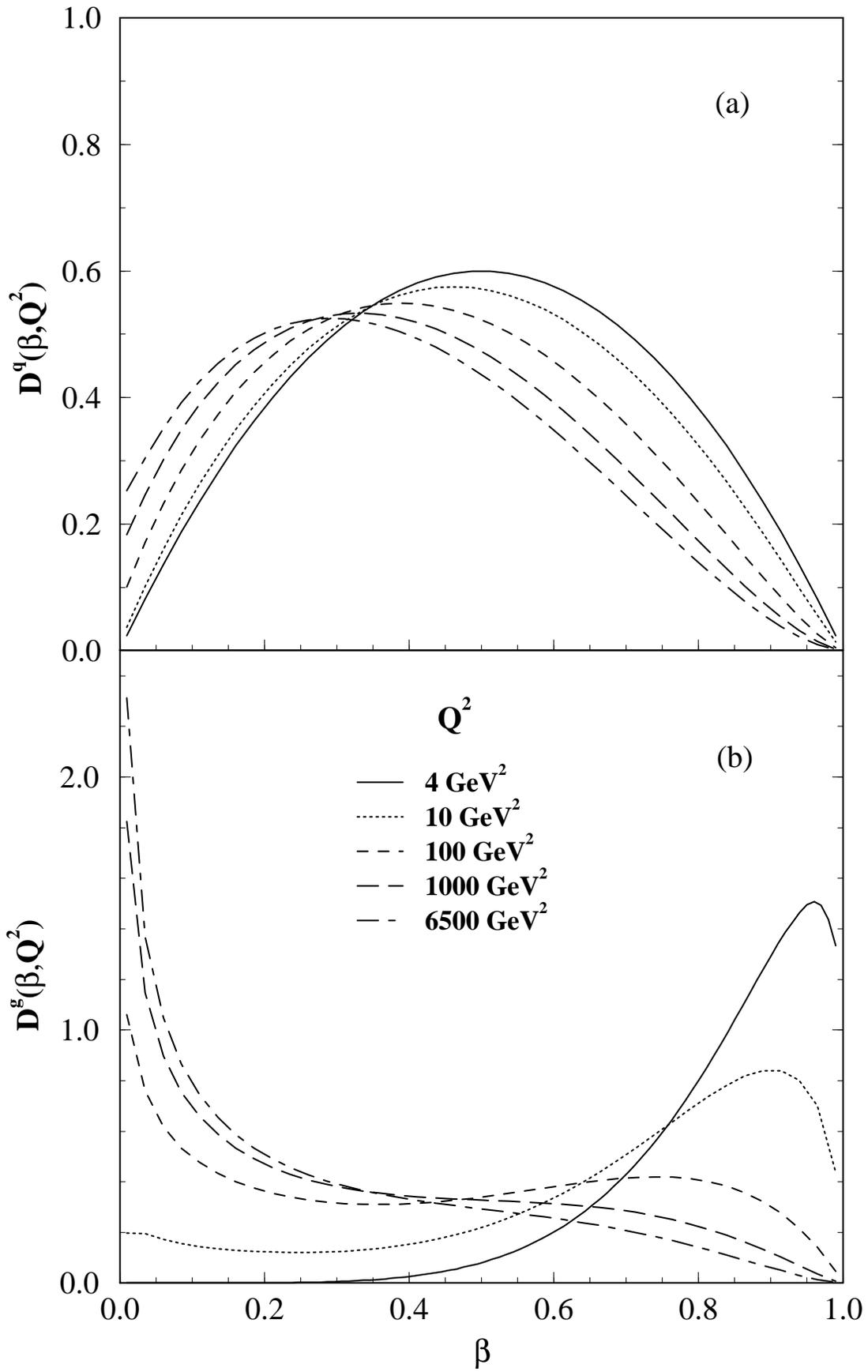
\noindent Figure 2: $Q^2$ evolution of the quark singlet (a) and gluon
(b) distributions obtained by the fitting procedure described in the 
text.

\vspace{1cm}

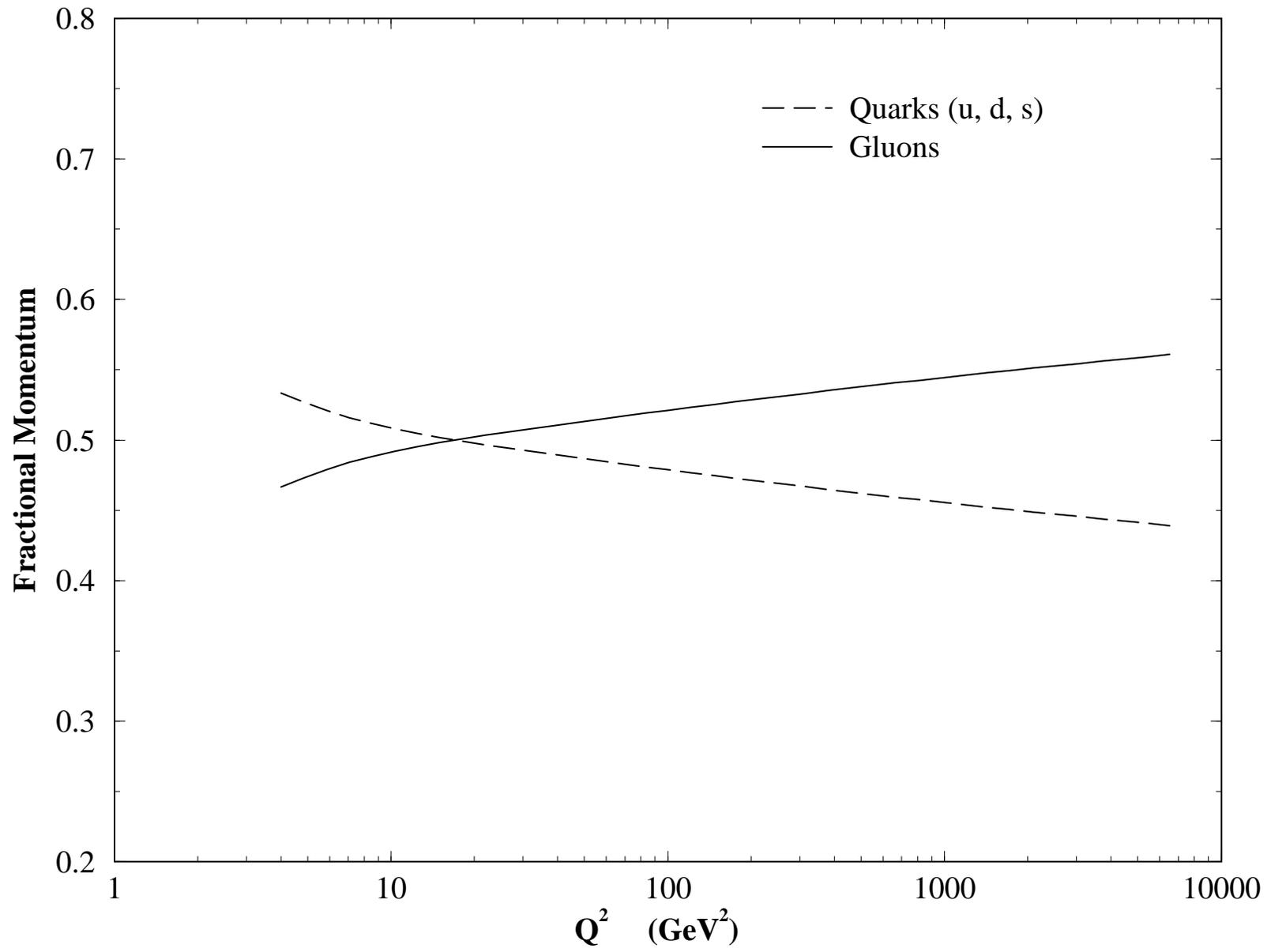
\noindent Figure 3:  $Q^2$ evolution of the fractions of the pomeron 
momentum carried by quarks and gluons as predicted from the 
parametrization resulted from the fit to the H1 data.

\end{document}